# A TUNABLE BINAURAL AUDIO TELEPRESENCE SYSTEM CAPABLE OF BALANCING IMMERSIVE AND ENHANCED MODES

*Yicheng Hsu[1] and Mingsian R. Bai[1,2]*

[1]Department of Power Mechanical Engineering, National Tsing Hua University, Taiwan
[2]Department of Electrical Engineering, National Tsing Hua University, Taiwan

## ABSTRACT

Binaural Audio Telepresence (BAT) aims to encode the acoustic scene at the far end into binaural signals for the user at the near end. BAT encompasses an immense range of applications that can vary between two extreme modes of Immersive BAT (I-BAT) and Enhanced BAT (E-BAT). With I-BAT, our goal is to preserve the full ambience as if we were at the far end, while with E-BAT, our goal is to enhance the far-end conversation with significantly improved speech quality and intelligibility. To this end, this paper presents a tunable BAT system to vary between these two AT modes with a desired application-specific balance. Microphone signals are converted into binaural signals with prescribed ambience factor. A novel Spatial COherence REpresentation (SCORE) is proposed as an input feature for model training so that the network remains robust to different array setups. Experimental results demonstrated the superior performance of the proposed BAT, even when the array configurations were not included in the training phase.

***Index Terms***— audio telepresence, binaural rendering, spatial feature, microphone array, deep learning

## 1. INTRODUCTION

Virtual reality is experiencing rapid growth, expanding beyond teleconferencing and broadcasting into diverse applications [1]. In the realm of audio virtual reality, audio telepresence (AT) attempts to bring the auditory scene of remote locations to nearby users via headsets or loudspeaker arrays. In this paper, we focus on Binaural Audio Telepresence (BAT) using headset rendering.

In a wide range of applications, it is desirable to have a tunable AT that can vary between either preserving the entire acoustic scene (filmmaking) or improving the signal quality by suppressing the noise, inference, and excessive reverberation (virtual videoconferencing). In a conventional BAT system, the acoustic scene is encoded using a spherical microphone array [2, 3], followed by Head-Related Transfer Function (HRTF) filtering [4]. This can be useful in the Immersive BAT (I-BAT) applications [5]-[7]. Conversely, Enhanced BAT (E-BAT) applications require signal enhancement [8]-[10] to mitigate the adverse effects of interference, noise, and reverberation. Many of these systems relied on ambisonic recording [2, 3] using spherical microphone arrays, which can add complexity to augmented reality applications that typically use irregularly shaped microphone arrays.

Rafaely et al. proposed a BAT system for a microphone-loaded glasses using a beamforming approach to identify directional sources, and then to produce the binaural signals via HRTF filtering, which can be considered as an E-BAT system [11]. Pulkki et al. used a subspace method to separate the directional and diffuse components for ensuing audio synthesis [12, 13]. These approaches rely on source counting and localization accuracy, and can fail in highly reverberant environments or when there are more sources than microphones. To address these challenges, we presented a BAT system by directly mapping array signals to binaural signals via a Deep Neural Network (DNN) termed Multichannel Deep Filtering (MDF) [14]. Despite the impressive rendering using MDF, this model can support either the I-BAT mode or the E-BAT mode. In addition, MDF was array-dependent, requiring retraining for different array configurations.

In this study, we propose a tunable BAT system capable of adapting to different array setups and balancing between the I-BAT and E-BAT modes. A novel Spatial COherence REpresentation (SCORE) spatial feature, which is robust to variations in array configuration, is used in network training. With SCORE, a Binaural Rendering Network (BRnet) based on DeepfFilterNet2 [15] is then used to convert the microphone array signals into binaural signals. The balance between the I-BAT and E-BAT modes is adjusted by a parameter in BRnet, facilitated by a Feature-wise Linear Modulation (FiLM) layer [16]. Several objective tests are performed to evaluate the proposed tunable BAT system over different array setups, using the MDF method [14] as a baseline.

The remainder of this paper is organized as follows. Section 2 introduces the signal model and the problem formulation. Section 3 presents the proposed tunable BAT system. Section 4 summarizes the experimental settings and results. Conclusions are given in section 5.

## 2. SIGNAL MODEL AND PROBLEM FORMULATION

Consider $M$ microphones capturing signals from speakers and interferers in a reverberant room. The signal obtained at the $m$th microphone can be represented in the STFT domain as

$$X^m(l,f) = \sum_{d=1}^{D} A_d^m(f) S_d(l,f) + \sum_{n=1}^{N} A_n^m(f) S_n(l,f) + V^m(l,f), \quad (1)$$

where $l$ and $f$ denote the time frame index and the frequency bin index, respectively, $A_d^m$ denotes the Acoustic Transfer function (ATF) between the $m$th microphone and the $d$th speaker, $A_n^m$ denotes the ATF between the $m$th microphone and the $n$th interferer, $S_d^m$ denotes the signal of the $d$th speaker, $S_n^m$ denotes the signal of the $n$th interferer, and $V^m$ denotes the sensor noise.

The BAT system aims to transform the microphone array signals into the desired binaural signals. The binaural output signals to render can be written as

$$\begin{aligned} Y_L(l,f) &= \sum_{d=1}^{D} H_{d,L}(f) A_{d\_clean}^m S_d(l,f) + \alpha R_L(l,f) \\ Y_R(l,f) &= \sum_{d=1}^{D} H_{d,R}(f) A_{d\_clean}^m S_d(l,f) + \alpha R_R(l,f) \end{aligned}, \quad (2)$$

where

$$\begin{aligned} R_L(l,f) &= \sum_{d=1}^{D} H_{d,L}(f) A_{d\_late}^m S_d(l,f) + \sum_{n=1}^{N} H_{n,L}(f) A_n^m S_n(l,f) \\ R_R(l,f) &= \sum_{d=1}^{D} H_{d,R}(f) A_{d\_late}^m S_d(l,f) + \sum_{n=1}^{N} H_{n,R}(f) A_n^m S_n(l,f) \end{aligned},$$

$A_{d\_clean}^m$ denotes the direct and early components of $A_d^m$, $A_{d\_late}^m$ denotes the late reverberation of $A_d^m$, $H_{d,L}$ and $H_{n,L}$ and $H_{n,R}$ denote the HRTFs from the $n$th interferer to the left and right ears, and $\alpha \in [0,1]$ is a tuning parameter to balance between the I-BAT ($\alpha=1$) and E-BAT ($\alpha=0$) modes. Note that the ambience includes the late reverberation of the directional sources and the interferers. To generate the $A_{d\_clean}^m$, the late reverberation of $A_d^m$ was exponentially attenuated, following the approach in [17]. The maximum decay of $A_{d\_clean}^m$ was shaped to T60 = 0.2 second in this study. In addition, $A_{d\_late}^m$ was generated by subtracting $A_d^m$ from the $A_{d\_clean}^m$.

## 3. PROPOSED TUNABLE BAT SYSTEM

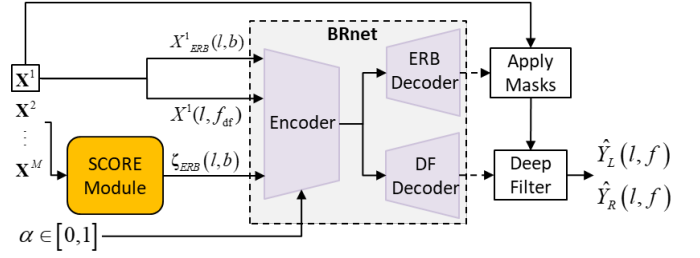

Fig. 1. The block diagram of the proposed BAT system.

This section presents a tunable BAT system for the user-specified balance of directional sound and ambience, as depicted in the block diagram in Fig. 1. The SCORE module provides spatial features to the subsequent BRnet, ensuring its adaptability to the array configuration, as will be detailed in Sec. 3.2. The tuning parameter $\alpha$ is required in the encoder to properly balance the I-BAT and E-BAT modes according to the application scenario.

### 3.1. The SCORE module

The short-term Relative Transfer Function (RTF) for each Time-Frequency (TF) bin is estimated by averaging $(2R+1)$ frames between the $m$th microphone and the reference first microphone

$$R^m(l,f) \equiv \frac{\sum_{n=l-R}^{l+R} X^m(l,f) X^{1*}(l,f)}{\sum_{n=l-R}^{l+R} X^1(l,f) X^{1*}(l,f)}, \quad (3)$$

where "*" is the complex conjugate operation. A critical step is to compute a "whitened" feature vector $\tilde{\mathbf{r}}(l,f)$ with the first entry (1) deleted corresponding to each TF bin:

$$\tilde{\mathbf{r}}(l,f) = \left[ \frac{R^2(l,f)}{|R^2(l,f)|}, \ldots, \frac{R^M(l,f)}{|R^M(l,f)|} \right]^T \in \mathbb{R}^{(M-1)\times 1}, \quad (4)$$

where $|\cdot|$ denotes the complex modulus.

With the deleted and whitened RTF defined above, the SCORE vector is constructed as follows

$$\zeta(l,f) = \frac{1}{M-1} \mathrm{Re}\{\mathbf{A}^H(f) \tilde{\mathbf{r}}(l,f)\} \in \mathbb{R}^{Q\times 1}, \quad (5)$$

where $\mathbf{A}(f) = [\mathbf{a}_1(f) \cdots \mathbf{a}_Q(f)] \in \mathbb{C}^{(M-1)\times Q}$ is the steering matrix composed of the deleted free-field plane-wave components at $Q$ uniformly spaced azimuthal angles, and $\mathrm{Re}\{\cdot\}$ denotes the element-wise real part. In this study, $Q$ is set to 12.

To reduce the computational complexity, SCORE is compressed using the Equivalent Rectangular Bandwidth (ERB) scale [18]:

$$\zeta_{ERB}(l,b) = \frac{1}{\pi_b} \sum_{f \in \{f_{b1},\ldots,f_{bF_b}\}} w_b(f) \zeta(l,f), \ b \in \{0,1,\ldots,B\}, \quad (6)$$

where $B$ is the total number of bands ($B = 32$ in this study), $w_b(f)$ and $F_b$ are the weight and the number of the frequency bins for the $b$th band, and $\pi$ is a normalization factor to ensure that each band is bounded in the range [-1, 1], defined as

$$\pi_b = \sum_f^{F_b} w_b(f). \quad (7)$$

### 3.2. The BRnet

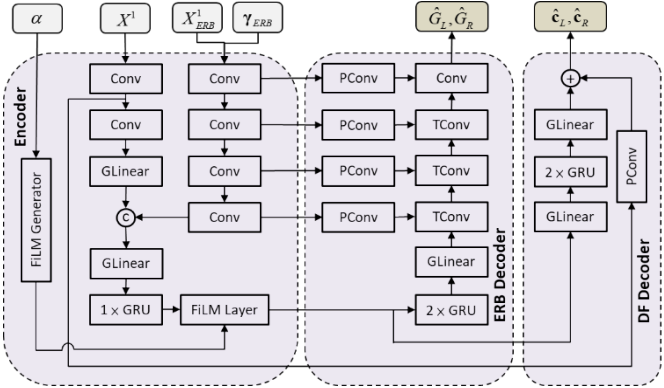

Fig. 2. The framework of BRnet. The Encoder captures both spatial and spectral information, which is used by the subsequent ERB decoder and DF decoder to generate masks and deep filter coefficients specific to the reference microphone.

The BRnet is based on DeepFilterNet2 designed according to human auditory characteristics [15], as shown in Fig.2. BRnet is also known for its impressive two-stage noise reduction performance. BRnet, consists of three main modules: the Encoder, the ERB decoder, and the Deep Filter (DF) decoder. Unlike the system proposed in [15], the proposed BRnet operates at 16 kHz and uses a 32-band ERB filterbank. A deep filter of order $N = 5$ is used only for the lowest 160 frequency bins, covering a bandwidth of 5 kHz. The ERB encoder has $(1+Q)$ input channels to accept the feature data. For binaural ERB mask generation, the ERB decoder provides two channels of binaural output. Similarly, two DF coefficients are predicted and applied separately to the left and right ears for phase restoration. Instead of directly combining the tuning parameter $\alpha$ with the embedding, this paper utilizes a FiLM layer [16] to perform a feature-wise affine transformation to facilitate the acquisition of conditional representations. The remaining parameters are the same as those in [15].

### 3.3. Training procedure and loss function

In this paper, each signal frame is 32 ms long and is processed using a 512-point FFT with a 256-point hop size. The Adam optimizer is used for training with a learning rate of 0.001. A gradient norm clipping of 3 is adopted. If the validation set loss shows no improvement for three consecutive epochs, the learning rate is halved. Frequency and time indices are omitted for brevity. The loss function used is the complex compressed mean square error suggested in [19], which includes both magnitude-sensitive and phase-sensitive terms

$$\mathcal{L} = \sum_{l,f} \left\| |\mathbf{Y}|^c - |\hat{\mathbf{Y}}|^c \right\|_F^2 + \sum_{l,f} \left\| |\mathbf{Y}|^c e^{j\angle \mathbf{Y}} - |\hat{\mathbf{Y}}|^c e^{j\angle \hat{\mathbf{Y}}} \right\|_F^2, \quad (8)$$

where $\mathbf{Y} = [\mathbf{Y}_L \ \mathbf{Y}_R]$ denotes binaural signal defined in Eq. (2) as the training target, $\hat{\mathbf{Y}} = [\hat{\mathbf{Y}}_L \ \hat{\mathbf{Y}}_R]$ denotes the estimated binaural signal, and the compression factor $c = 0.3$ as suggested in [18]. Both magnitude compression and angle extraction are element-wise operations.

### 3.4. Training the model for the tunable BAT

This paper presents a simple yet effective training approach aimed at achieving a pre-specified balance between I-BAT and E-BAT modes. The tunable parameter $\alpha$ randomly selected from {0.0, 0.3, 0.5, 0.7, 1.0} during each training iteration. Using a supervised learning procedure, the model learns to fine-tune the ambience component by projecting the embedding vector over the FiLM layer according to the desired $\alpha$.

## 4. EXPERIMENTAL STUDY

### 4.1. Data preparation

Clean speech signals containing utterances from 921 speakers and 40 speakers were selected from the train-clean-360 and test-clean subsets of the LibriSpeech corpus [20] for training and testing the proposed model. The interferers were generated using the music signals form the MTG-Jamendo dataset [21]. A sample rate of 16 kHz was used for all experiments. A total of 50,000 samples were used for training and 5,000 samples for validation. Noise mixtures, edited in five-second clips, were created by mixing the speech signals from two speakers with an interferer at various Signal-to-Interference Ratios (SIRs) of 5, 10, and 15 dB. The two speaker signals were set to the same pressure level. Gaussian white sensor noise was added with Signal-to-Noise Ratio (SNR) of 20, 25, and 30 dB. Figure 3 illustrates the experimental settings used in this study. The speech and music signals were convolved with the Room Impulse Responses (RIRs) in different directions to create the microphone signals. A five-element array, G1, used for training and validation is shown in Fig. 4. RIRs were simulated using the image source method [22] with different reverberation times (T60), 0.3, 0.4, 0.5, and 0.6 seconds. For binaural signals, the HTRFs were also convolved with the corresponding sources. HRTFs were selected from the SADIE II database provided by the University of York [23].

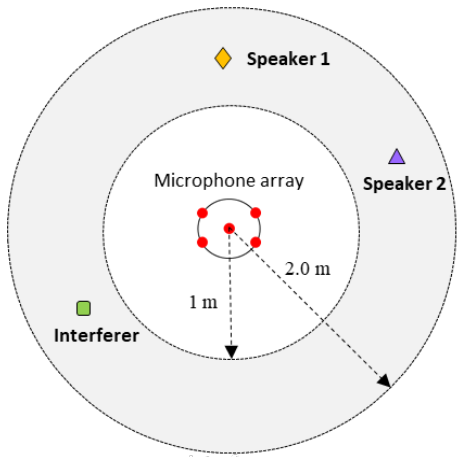

Fig. 3. Experimental setup. Sources are randomly positioned within a ring sector. A minimum angular separation of 30° between any two sources.

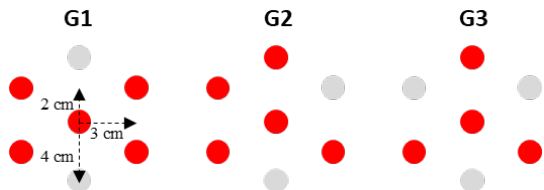

Fig. 4. The microphone array setups used in the experiments to investigate the effects of different array configurations.

**4.2. Baseline and implementation details**

A learning-based approach, MDF [14], is used as a baseline in this study. The training procedure and the loss function are the same as those described in Sec. 3.3. However, the MDF network (MDFnet) can support either I-BAT or E-BAT mode, but not both, let alone balanced mode. Therefore, two MDFnets were employed as baselines for the proposed tunable BAT: one called I-MDF, where $\alpha = 1$ for I-BAT, and E-MDF, where $\alpha = 0$ specifically for E-BAT.

**4.3. Results**

To evaluate the robustness of the proposed system, three 200-sample datasets are used in the testing phase. These datasets were generated using measured RIRs from the Rotated Circular Array dataset provided by Tampere University [24], with different array setups (G1, G2, G3) as shown in Fig. 4. In this study, three objective indices: magnitude-weighted Interaural Phase Difference error (mw-IPDe), magnitude-weighted Interaural Level Difference error (mw-ILDe), and modified Scale-Invariant Signal-to-Distortion Ratio (mSI-SDR) suggested in [25] are used to evaluate the BAT performance.

Table I summarizes the I-BAT, I/E-BAT, and E-BAT performance achieved by the proposed tunable BAT system when $\alpha$ = 1, 0.5, and 0, respectively, compared to the MDF baseline [14]. The performance of MDF is degraded significantly by the unseen array geometries (G2). MDF cannot work for G3 due to its fixed input dimension. In contrast, the proposed BAT system trained with the SCORE feature showed superior performance in I-BAT and E-BAT, even for unseen array geometries (G2 and G3) and sensor counts (G3). Note that MDF requires to train two different MDFnets (I-MDF and E-MDF) with $\alpha$ = 1 and 0 corresponding to the I-BAT and E-BAT modes, respectively. In contrast, the proposed tunable BAT system has a unique network architecture by adjusting the tuning parameter in BRnet. This results in a remarkable reduction in complexity, such that BRnet (2.5 M) requires only half the parameter size of MDFnet (6.36 M) [14], without significantly compromising performance and robustness.

Table I. Comparison of BAT performance in terms of mw-IPDe, mw-ILDe, and mSI-SDR for different array configurations.

| | Array setup | Method | mw-IPDe | mw-ILDe | mSI-SDR |
|---|---|---|---|---|---|
| I-BAT ($\alpha$ = 1) | G1 | I-MDF | 0.60 | 7.64 | 1.85 |
| | | Proposed | **0.49** | **5.50** | **5.20** |
| | G2 | I-MDF | 1.10 | 9.39 | -8.01 |
| | | Proposed | **0.58** | **5.60** | **4.72** |
| | G3 | I-MDF | N/A | N/A | N/A |
| | | Proposed | **0.65** | **5.81** | **3.88** |
| I/E-BAT ($\alpha$ = 0.5) | G1 | MDF | N/A | N/A | N/A |
| | | Proposed | **0.46** | **5.33** | **5.26** |
| | G2 | MDF | N/A | N/A | N/A |
| | | Proposed | **0.55** | **5.50** | **4.71** |
| | G3 | MDF | N/A | N/A | N/A |
| | | Proposed | **0.62** | **5.71** | **3.86** |
| E-BAT ($\alpha$ = 0) | G1 | E-MDF | 0.59 | 7.86 | -0.21 |
| | | Proposed | **0.47** | **5.33** | **3.49** |
| | G2 | E-MDF | 1.20 | 10.15 | -10.99 |
| | | Proposed | **0.55** | **5.53** | **3.12** |
| | G3 | E-MDF | N/A | N/A | N/A |
| | | Proposed | **0.60** | **5.79** | **2.28** |

**5. CONCLUSIONS**

In this study, a tunable BAT system has been proposed to convert microphone array signals into the desired binaural outputs with prescribed balance of directional and diffuse components. The experimental results showed that the proposed SCORE is a spatial feature robust to unseen array configurations. BRnet has shown remarkable performance in with significantly reduced complexity compared to MDFnet.